\documentclass[aps,prl,twocolumn,groupedaddress,english]{revtex4-1}
\usepackage[utf8]{inputenc}
\usepackage[english]{babel}
\usepackage{amssymb,amsmath}
\usepackage{graphicx}
\usepackage[colorlinks=true, urlcolor=blue, citecolor=blue, filecolor=blue, linkcolor=blue]{hyperref}
\usepackage{siunitx}
\usepackage{bigstrut}
\usepackage{tabularx}
\usepackage{braket}
\usepackage{longtable}
\usepackage{comment}

\def \IPRaunhardtJCP {34301.2(3)}
\def \LowRotIntIonRaunhardtJCP {70.943(7)}
\def \LowRotIntMetastableRaunhardtJCP {}

\def \IPSprecherJCP {34 301.211(10)}
\def \LowRotIntIonSprecherJCP {70.937(3)}

\def \IPJansenPRL {34301.2059(14)}
\def \LowRotIntMetastableJansenPRL {75.8137(4)}
\def \LowRotIntIonJansenPRL {70.9380(6)}

\def \IPFinalPrecision {1.3}

\def \IPFinalValue {34301.207002(23)}
\def \IPFinalValueAllUncertainties {34301.20700(4)}
\def \LowRotIntMetastableFinalValue {75.812953(20)}
\def \LowRotIntIonFinalValue {70.937589(23)}
\def \LowRotIntIonFinalValueAllUncertainties {70.93759(6)}
\def \TotalSystIP {0.000037}
\def \TotalSystInterval {0.000060}

\def \LowRotIntMetastableFocsaJMS	 {75.8129(3)}

\def \IPGinterJMS  {34301.22(10)}
\def \LowRotIntIonGinterJMS {70.99(14)}

\def \LowRotIntIonTungJCP {70.936} 

\def \LowRotIntIonMatyusJCP {70.938}

\newcommand{\plussign}{\raisebox{.4\height}{\scalebox{.6}{+}}}

\def \FirstOrderDopplerUncertaintyCombPaper {460}
\def \ReproducibilityUncertaintyCombPaper {700} 

\begin{document}

\title{Precision measurements in few-electron molecules:\\ The ionization energy of metastable $\mathbf{^4}$He$\mathbf{{_2}}$ and the first rotational interval of $\mathbf{^4}$He$\mathbf{{_2}^+}$}
\author{Luca Semeria}
\author{Paul Jansen}
\email[]{paul.jansen@phys.chem.ethz.ch}
\author{Gian-Marco Camenisch}
\author{Federico Mellini}
\author{Hansj\"urg Schmutz}
\author{Fr\'ed\'eric Merkt}

\affiliation{Laboratory of Physical Chemistry, ETH Zurich, CH-8093 Zurich, Switzerland}

\date{\today}

\begin{abstract}
Molecular helium represents a benchmark system for testing \textit{ab initio} calculations on few-electron molecules. We report on the determination of the adiabatic ionization energy of the $a\,^3\Sigma_u^+$ state of He$_2$, corresponding to the energy interval between the $a\,^3\Sigma_u^+$ ($v''=0$, $N''=1$) state of He$_2$ and the $X^+\,^2\Sigma_u^+$ ($v^+=0$, $N^+=1$) state of He${_2}^+$, and of the lowest rotational interval of He${_2}^+$. 
These measurements rely on the excitation of metastable He$_2$ molecules to high Rydberg states using frequency-comb-calibrated continuous-wave UV radiation in a counter-propagating-laser-beam setup. The observed Rydberg states were extrapolated to their series limit using multichannel quantum-defect theory. 
The ionization energy of He$_2$ ($a\,^3\Sigma_u^+$) and the lowest rotational interval of He${_2}^+$ ($X^+\,^2\Sigma_u^+$) are \IPFinalValue$\pm\TotalSystIP_{\mathrm{sys}}$\,cm$^{-1}$ and \LowRotIntIonFinalValue$\pm\TotalSystInterval_{\mathrm{sys}}$\,cm$^{-1}$, respectively.

\end{abstract}


\maketitle


The comparison of the results of precision spectroscopic measurements in few-electron atoms and molecules with the results of \emph{ab initio} calculations challenges theoreticians and experimentalists and forces them to constantly improve their respective methodologies. In recent years, this symbiotic relation between theory and experiment has opened a route to test the Standard Model of particle physics and extensions thereof \cite{steimle2014,altmann2018,safronova2018,karshenboim2018} and to determine the values of fundamental constants \cite{karr2016,biesheuvel2016,biesheuvel2017,alighanbari2018,kato2018}. The 2019 revision of the International System of Units (SI) fixed the Boltzmann ($k_\mathrm{B}$) and Avogadro ($N_\mathrm{A}$) constants \cite{newell2019}, enabling a primary pressure standard that is directly traceable via the SI to the density of a gas. The number density of a gas can be determined via a measurement of the refractive index or the dielectric constant. Both methods rely on the precise value of the polarizability of the gas, which is known with sufficient accuracy only for atomic helium from recent \emph{ab initio} calculations~\cite{puchalski2016a,puchalski2020}. An accurate determination of the level energies of He${_2}^+$ might provide a way to test the polarizability calculations of atomic helium via the long-range part of the molecular potential
\begin{equation}
 V(R)=-\frac{\alpha_0}{2R^4} +\alpha^2 \frac{W_4}{R^4} +\mathcal{O}(R^{-6})\text{ for }R\rightarrow \infty\label{eq:VLR_Comb_Paper},
\end{equation}
where $\alpha_0=1.38376077(14)$\,$a_0^3$ is the calculated static polarizability of He (1\,$^1\text{S}_0$) \cite{puchalski2016a}, $\alpha$ is the fine-structure constant and $\alpha^2 W_4 R^{-4}$ is the relativistic correction from the Breit interaction~\cite{meath1966}. Terms in $R^{-6}$, which can be calculated accurately, ought to be included in a fit based on Eq.~\eqref{eq:VLR_Comb_Paper}. Equation~\eqref{eq:VLR_Comb_Paper} becomes a good approximation of the internuclear potential for $R>R_\text{LR}\approx 4\,a_0$, with $R_\text{LR}$ being the Le Roy radius~\cite{leroy1974}, implying that the highest vibrational levels of He${_2}^+$ are the most sensitive to the value of $\alpha_0$. 
Model calculations indicate that an uncertainty of $1.4\times 10^{-6}$\,$a_0^3$ might shift the absolute positions of these levels by up to 0.2\,MHz. 

Transition frequencies between low-lying rotational levels of the vibronic ground state of $^3$He$^4$He$^+$ were determined with a precision of 18\,MHz \cite{yu1987} but are not sensitive to $\alpha_0$. 
The frequencies of transitions between highly vibrationally excited bound levels of the $X^+$\,$^2\Sigma_u^+$ and weakly bound $A^+$\,$^2\Sigma_g^+$ levels of He${_2}^+$ have been measured with a 1\,MHz precision \cite{carrington1995} but they are not sensitive to $\alpha_0$ either because the effects of the long-range interactions largely cancel out in the energy differences. Experimental data on a broader range of levels of He${_2}^+$  are thus required. 

We present the results of a determination of the adiabatic ionization energy of metastable He$_2$ in its \textit{a}\,$^3\Sigma_u^+$ state (He$_2^*$ hereafter) and the first rotational interval of He${_2}^+$ in its vibronic $X^+$\,$^2\Sigma_u^+$\,$(v^+=0)$ ground state (He${_2}^+$ hereafter) at a relative accuracy of $10^{-9}$ and $8.5\times10^{-7}$, respectively. These results improve our previous results on this system by more than an order of magnitude \cite{jansen2015,semeria2016} and approach the level of accuracy required by the metrological application described above.
They also serve as benchmark data for \textit{ab initio} calculations of three- and four-electron systems \cite{tung2012,matyus2018}.

\begin{figure}[!ht]
\centering
\includegraphics[width=1\columnwidth]{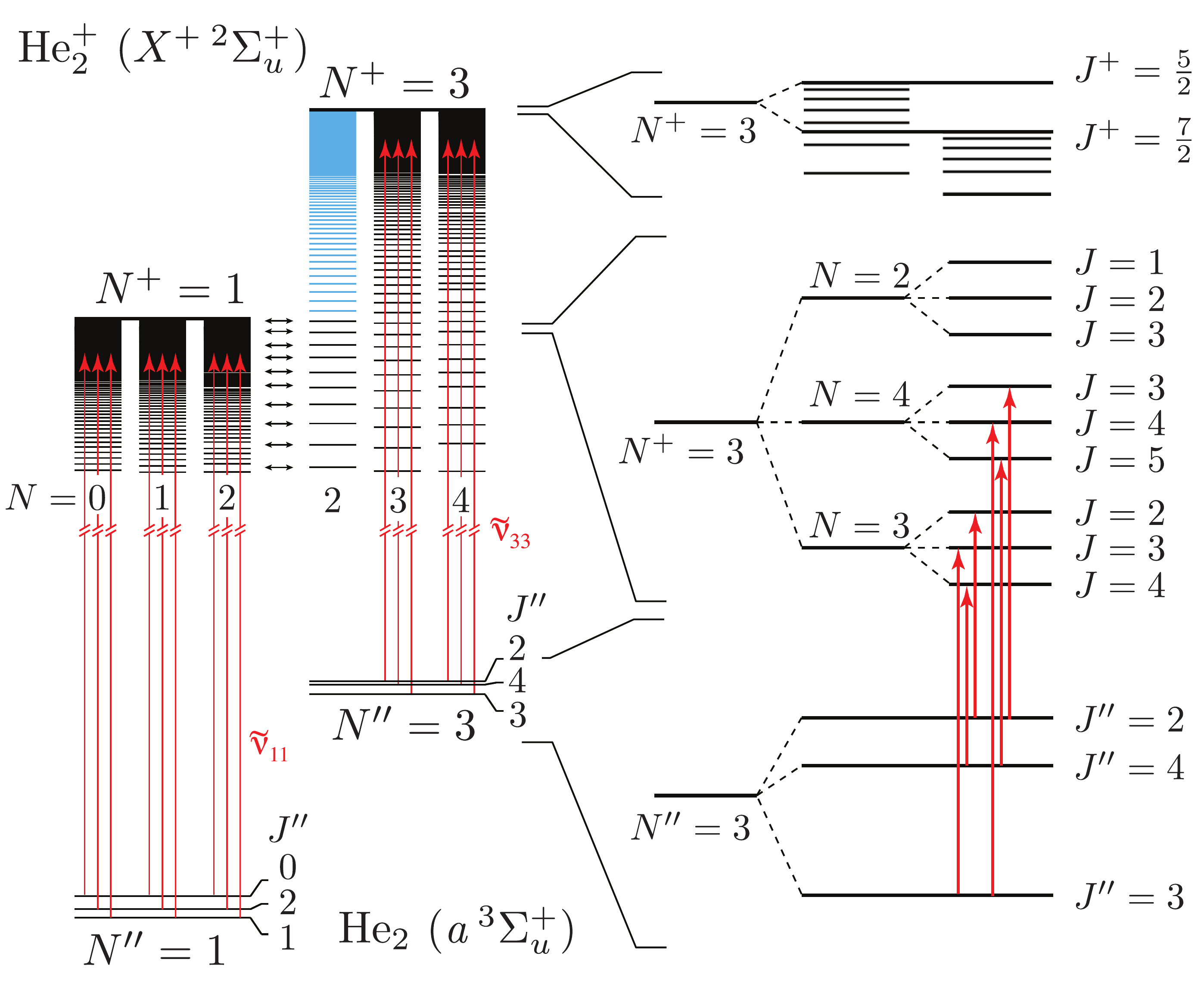}
\caption{Energy-level diagram (not to scale) showing the lowest rovibrational states of He$_2^*$ and $n\text{p}$ Rydberg series converging on the corresponding rovibrational states of He${_2}^+$. Relevant allowed transitions to $n$p Rydberg states are indicated by full arrows. Autoionizing levels are marked in blue (see text for details).
\label{fig:EnergyDiagram_Comb_Paper}}
\end{figure}

Our approach consists of measuring Rydberg series of He$_2$ from the metastable \textit{a}\,$^3\Sigma_u^+$ state of He$_2$ and extrapolating them to the series limits.
The two lowest rotational levels of He$_2^*$ and the $n$p Rydberg series converging on the corresponding rotational levels of He${_2}^+$ are schematically depicted in Fig.~\ref{fig:EnergyDiagram_Comb_Paper}. 
Only odd rotational levels $N''$ and $N^+$ are allowed in He$_2^*$ and He${_2}^+$, respectively, because $^4$He$^{2\plussign}$ is a boson.
We use double-primed symbols, unprimed symbols, and a superscript ``$+$'' to designate the quantum numbers of He$_2^*$, He$_2$ Rydberg states, and He${_2}^+$, respectively. The level structure of He$_2^*$ is adequately described using Hund's angular-momentum coupling case (b), i.e., the total angular momentum excluding spin $\vec{N}''$ couples to the total electron spin $\vec{S}''$ to form the total angular momentum $\vec{J}''$. The splittings between the resulting fine-structure components with $J''=N'', N''\pm 1$ are known very accurately from radio-frequency measurements \cite{lichten1974,semeria2018}. The level structure of He${_2}^+$ is also best described using Hund's angular-momentum coupling case (b). The rotational levels of He${_2}^+$ are split into two spin-rotation components with $J^+=N^+\pm\tfrac{1}{2}$ and the splittings for rotational states with $N^+\le 27$ were recently measured, with an accuracy of 150\,kHz for $N^+=1$ \cite{jansen2018}.  

Rydberg states of He$_2$ with principal quantum numbers $n\lesssim 200$ are well described in Hund's case (d) coupling. The rotation of the ionic core excluding spin $\vec{N}^+$ couples to $\vec{\ell}$ to give $\vec{N}$, which couples to $\vec{S}$ to give the total angular momentum of the Rydberg state $\vec{J}$. The triplet ($S=1$) $n\text{p}$ Rydberg states of He$_2$ are split in nine fine-structure components \cite{jansen2018}. For $n\gtrsim 200$, Hund's coupling case (e[b]) applies, in which singlet and triplet states are mixed (See Ref.~\cite{jansen2018} for details). The energy-level structures resulting from these coupling schemes are schematically depicted on the right-hand side of Fig.~\ref{fig:EnergyDiagram_Comb_Paper}. 

Because the transition-dipole-moment operator does not act on spins, transitions from He$_2^*$ to $n$p Rydberg states are governed by the selection rule $\Delta J=J-J''=N-N''=\Delta N$ \cite{jansen2018}. The observed Rydberg-transition wavenumbers are extrapolated to their series limit using multichannel quantum-defect theory (MQDT) \cite{jungen2011}, yielding the adiabatic ionization energy of He$_2^*$ and the first rotational interval of He${_2}^+$. The Rydberg transitions observed in our spectra connect the Hund's case (b) levels of He$_2^*$ to the Hund's case (d) Rydberg levels and can be labeled as $N''_{J''}\rightarrow n\text{p}N^+_{N,J}$. Here we also report fine-structure-free values for the transition wavenumbers, in which case the subscript labels $J''$ and $J$ are omitted. 

The measurements were carried out using the experimental apparatus described in Ref.~\cite{jansen2018}, specifically adapted to (i) record first-order-Doppler-corrected single-photon transitions \cite{beyer2016,cheng2018,holsch2019} and (ii) calibrate the transition frequencies using a frequency comb \cite{deiglmayr2016}.

Pulsed supersonic beams of He$_2^*$ molecules are produced in a discharge of pure helium~\cite{raunhardt2008}. Beam velocities of about 1000\,m/s and 500\,m/s are obtained by cooling the valve body to 77\,K and 10\,K, respectively \cite{motsch2014}. A 3-mm skimmer selects the most intense part of the molecular beam, which intersects the laser radiation at near right angles after about 1\,m of free flight. 
Transitions to Rydberg states with $n\gtrsim 30$ are detected by applying a pulsed electric field of 750\,V/cm across a cylindrically symmetric stack of electrodes. The pulsed ionization field also serves to accelerate the produced He${_2}^+$ ions toward an off-axis microchannel-plate (MCP) detector. A spectrum is recorded by monitoring the ion yield as a function of the laser frequency.

The narrow-bandwidth ($<1$\,MHz) UV radiation used for the excitation of He$_2^*$ to $n$p Rydberg states is produced by frequency doubling the output of a cw ring dye laser tunable around 580\,nm with a beta-barium borate (BBO) crystal mounted in a feedback-stabilized external cavity. 
The Doppler-corrected transition frequencies are measured by retroreflecting the laser beam after it passes the interaction zone, thereby generating two Doppler-shifted components in the spectrum. Their geometric mean corresponds to the first-order-Doppler-free transition frequency in case of a perfect overlap between the two counter-propagating beams.
Care was taken to minimize the waist of the laser beam, by means of a telescope, at the position corresponding to the surface of the backreflecting mirror, which was used to overlap the forward- and backward-propagating beams over a path length of about 10\,m. The systematic uncertainty associated with the difficulty of achieving a perfect overlap of the two beams (see Ref.~\cite{beyer2016} for details) effectively transformed into a statistical uncertainty of \FirstOrderDopplerUncertaintyCombPaper\,kHz when repeating the measurements multiple times after complete realignment of the laser beams.

Absolute calibration of the fundamental laser frequency relies on the use of an optical frequency comb (Menlo Systems FC1500-250-ULN), referenced to a GPS-disciplined Rb oscillator. 
The Rb oscillator (Stanford Research Systems FS725) has a stability of $2\times10^{-11}$ over the time (1\,s) required to record one data point and the GPS receiver (Spectrum Instruments TM-4) has a specified long-term stability of $10^{-12}$ \cite{deiglmayr2016}.
The beat signal measured with the photodiode is filtered by the frequency-comb electronics and is compared with a passive frequency discriminator inside a home-built electronic unit to generate an error signal, as described in Ref.~\cite{beyer2018}. This signal is used to control the reference cavity of the laser \cite{beyer2018} and to ensure, by means
of a proportional-integral-derivative (PID) feedback-controlled loop, that the beat frequency is locked at 60\,MHz \cite{ritt2004}.
The frequency of the laser follows from $f_\mathrm{L}=n_{\mathrm{c}} f_\mathrm{r} \pm 2f_\mathrm{CEO} \pm f_\mathrm{beat}$, where $f_\mathrm{CEO}$ and $f_\mathrm{r}$ are the carrier-envelope-offset and repetition frequency of the frequency comb, respectively, and $n_{\mathrm{c}}$ is the mode number that generates the beat signal, as determined using a wavemeter.

A small dc electric potential is applied to the stack to compensate the stray electric field along the symmetry axis. Four pin electrodes are employed to compensate the stray field in the plane perpendicular to the symmetry axis, resulting in an overall field compensation of better than 1\,mV/cm \cite{beyer2018}, corresponding to a measured frequency shift of less than 200\,kHz for Rydberg states of $n \approx 100$. The interaction region is enclosed in two concentric mumetal shields to suppress stray magnetic fields to better than 1\,mG \cite{semeria2016}.

\begin{figure}[!hb]
\centering
\includegraphics[width=1\columnwidth]{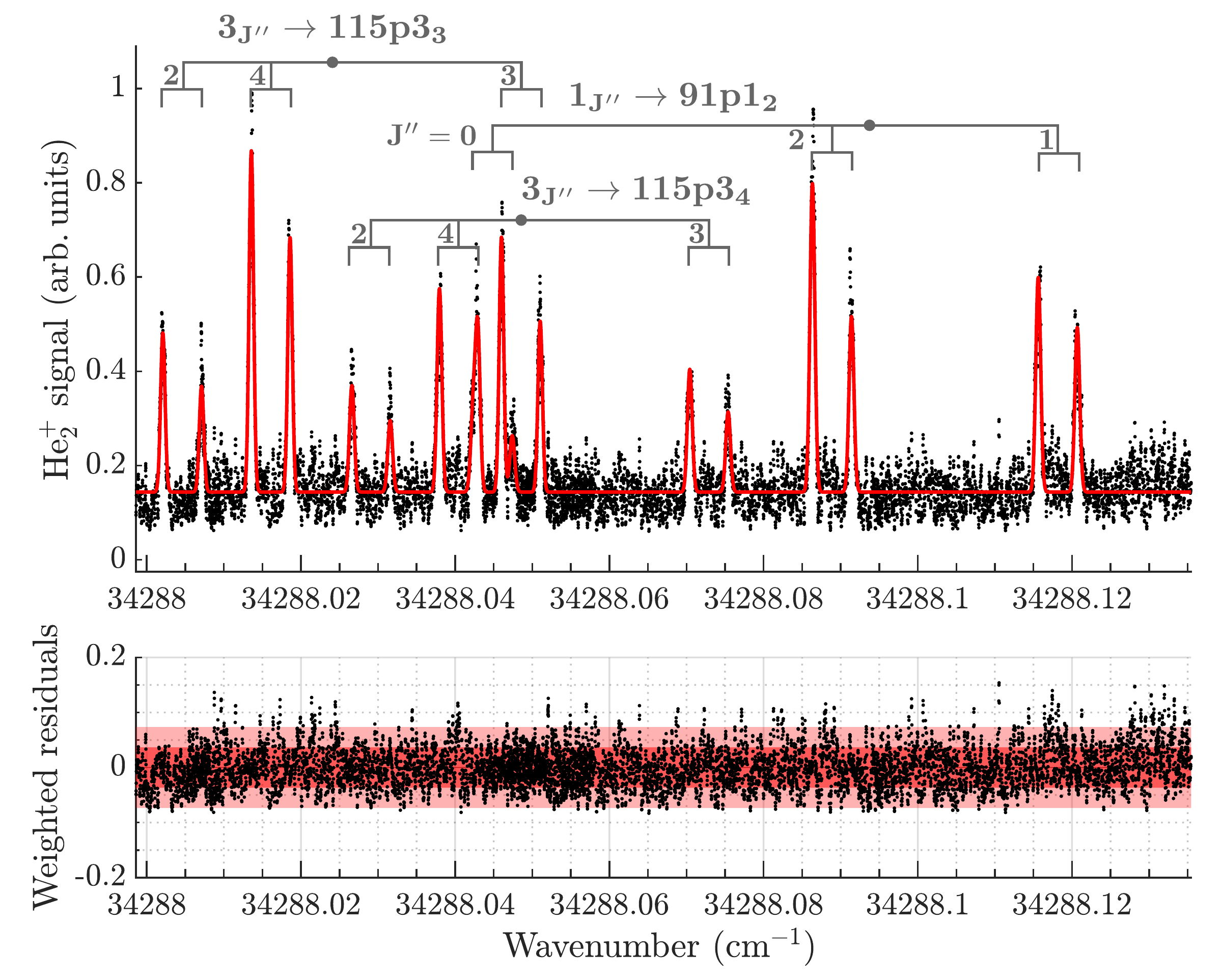}
\caption{Upper panel: Experimental data (black dots) and weighted-fit model (red curve) of transitions to Rydberg states converging on the $v^+=0$, $N^+=1,3$ levels of He${_2}^+$. The assignment is indicated by the grey bars, on which the small circle denotes the centre of gravity of the transitions of each triplet. Lower panel: Weighted residuals. The light and dark-shaded areas correspond to one and two standard deviations, respectively.
\label{fig:spectrum_with_fit_Comb_Paper}}
\end{figure}
Figure~\ref{fig:spectrum_with_fit_Comb_Paper} shows a Rydberg spectrum in the region of the overlapping $1\rightarrow 91$p$1_2$ and $3\rightarrow 115$p$3_{3,4}$ transitions.
The spectrum was recorded at a valve temperature of 77\,K, resulting in a Doppler-limited linewidth of 21\,MHz (full width at half maximum). 
Each transition appears as a pair of Doppler components generated by the counter-propagating laser beams. 
The Doppler pairs can be grouped in triplets reflecting primarily the fine-structure splittings in the metastable state \cite{jansen2018}, and can be determined separately using the $\Delta J=\Delta N$ selection rule (see Fig.~\ref{fig:EnergyDiagram_Comb_Paper} and Ref.~\cite{jansen2018}). 
The lineshape model used to extract the fine-structure-free transition wavenumbers is the sum of three Gaussian pairs. The frequency spacing between the pair of peaks corresponds to twice the first-order Doppler shift and is represented by a single parameter in the fit. The width of all Gaussians is also accounted for by a single parameter in the fit. The remaining fit parameters are a constant background offset, the separations between the first-order Doppler-corrected transition frequencies, and the individual peak intensities. 
The Doppler-corrected transition frequencies of each triplet are converted into their fine-structure-free center-of-gravity (c.o.g.) positions using the $2J^{\prime\prime}+1$ statistical weight of each fine-structure transition and are listed in Table~\ref{tab:transitions_table_Comb_Paper}. 
The reported fit values were obtained in a two-step approach to account for the Poissonian statistics of the ion detection \cite{naus2001}. An initial unweighted fit was used to estimate the Poissonian variance that, along with the variance of the intrinsic noise, was used to determine the statistical weights used in the fits.

\begingroup
\squeezetable
\begin{table}[!ht]
\caption{\label{tab:transitions_table_Comb_Paper}
Observed transitions from the $a\,^3\Sigma_u^+$ $(\nu''=0, N''=1,3)$ state of $^4$He$_2$ to the $n$p$N^+_N$ Rydberg states belonging to series converging to the $X^+\,^2\Sigma_u^+$ $(\nu^+=0, N^+=1,3)$ states of $^4$He${_2}^+$ and comparison with the results of MQDT calculations. The symbol $\Delta = (\text{obs.}-\text{calc.})$ represents the difference between observed and calculated line positions. All values are given in cm$^{-1}$.}
\begin{tabular*}{\columnwidth}{
s[table-unit-alignment = right,table-format=3]
@{\extracolsep{\fill}} 
S[table-format=6.8]
S[table-format=4.4]
S[table-format=6.8]
S[table-format=4.4]
}
\hline\hline
  & \multicolumn{2}{c}{$1\rightarrow n\text{p}1_1$}& 
\multicolumn{2}{c}{$1\rightarrow n\text{p}1_2$} \\
\cline{2-3}\cline{4-5}
\\[-1.8ex]
\multicolumn{1}{c}{$n$}&
\multicolumn{1}{c}{$\tilde{\nu}_\text{obs}\,(\sigma_{\tilde{\nu}_\text{obs}})$}&
\multicolumn{1}{c}{$\Delta\times10^{6}$}& 
\multicolumn{1}{c}{$\tilde{\nu}_\text{obs}\,(\sigma_{\tilde{\nu}_\text{obs}})$}&
\multicolumn{1}{c}{$\Delta\times10^{6}$}
\\[-1.8ex]
\\ \hline
\\[-1.8ex]
 70 &				&			&	34278.903916(6)	&	-30	\\
 75 &		34281.662500(8)		&	34		&		34281.687568(10)		&	24 	\\
 78 &		34283.138307(9)		&	13		&				&		\\
 81 &				&			&		34284.523515(7)		&	7 	\\
 83 &				&			&		34285.298807(20)		&	10	\\
  85 &	34285.994074(15)			&	10		&	34286.025503(10)			&	22	\\
 89 &		34287.331507(9)		&	-32		&		34287.322442(6)		&	39  	\\
 91 &				&			&		34288.093742(8)		&	-40 	\\
 94 &		34288.769663(9)		&	 -2		&		34288.828712(8)		&	-1 	\\
  97&				&			&		34289.563480(23)		&	10 	\\
\\[-1.8ex]
\hline\hline
& \multicolumn{2}{c}{$3\rightarrow n\text{p}3_3$}& 
\multicolumn{2}{c}{$3\rightarrow n\text{p}3_4$} \\
\cline{2-3}\cline{4-5}
\\[-1.8ex]
\multicolumn{1}{l}{$n$}&
\multicolumn{1}{c}{$\tilde{\nu}_\text{obs}\,(\sigma_{\tilde{\nu}_\text{obs}})$}&
\multicolumn{1}{c}{$\Delta\times10^{6}$}& 
\multicolumn{1}{c}{$\tilde{\nu}_\text{obs}\,(\sigma_{\tilde{\nu}_\text{obs}})$}&
\multicolumn{1}{c}{$\Delta\times10^{6}$}
\\[-1.8ex]
\\ \hline
\\[-1.8ex]
 103&	34285.974192(8)			&	-12		&				&		\\
 115&	34288.024156(9)		&	-11		&		34288.048567(12)		&	14	\\
 117&	34288.305868(8)		&	-23		&		34288.327945(10)			&	29	\\
 121&	34288.827917(12)		&	-32		&	34288.846249(10)		&	48	\\
\\[-1.8ex]
\hline\hline
\end{tabular*}
\end{table}
\endgroup

\begin{figure}[!ht]
\centering
\includegraphics[width=1\columnwidth]{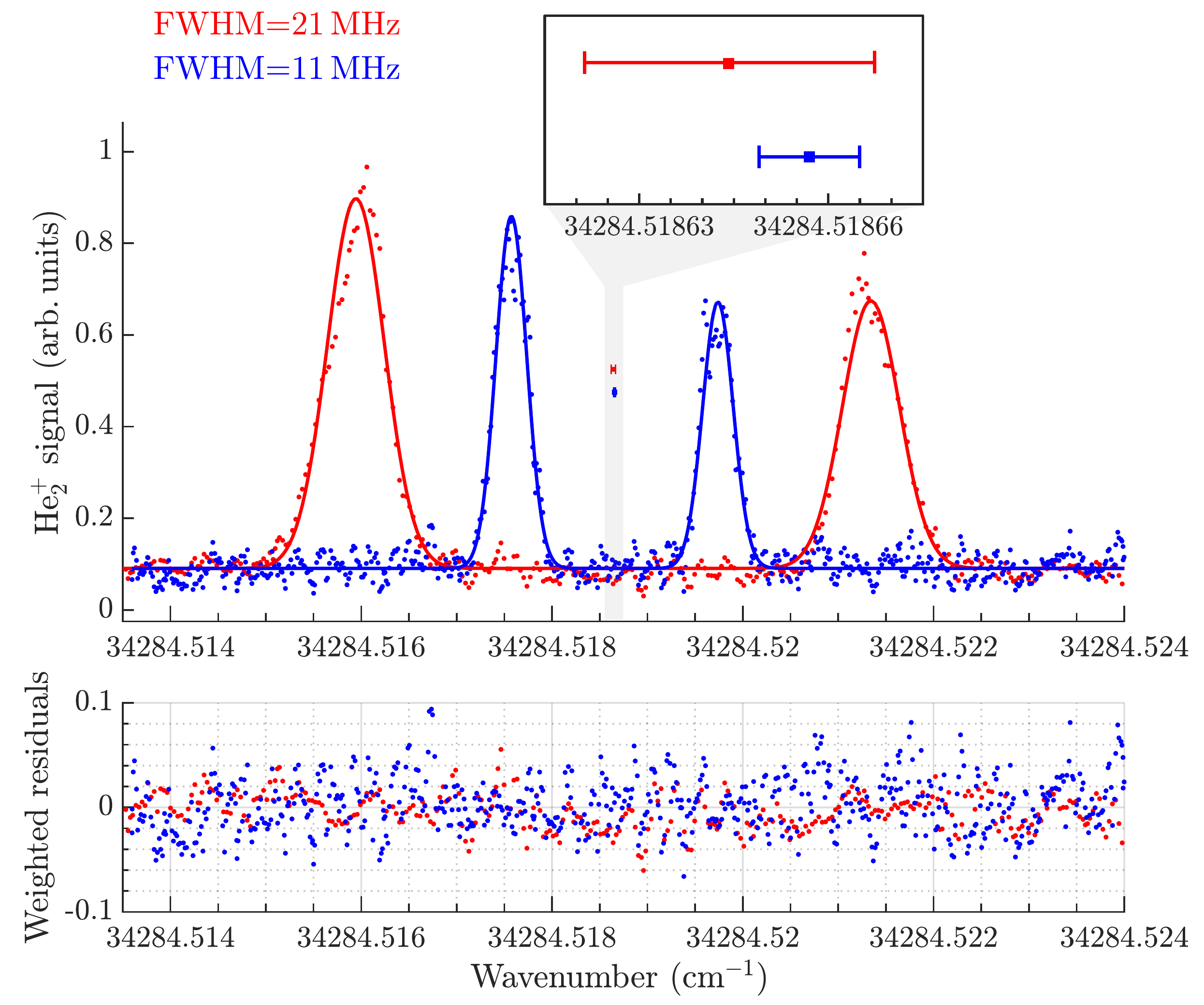}
\caption{Comparison of the fitted $1_2\rightarrow 81$p$1_2$, Rydberg spectra recorded at $77$\,K (red trace) and $10$\,K (blue trace), resulting in a FWHM of 21\,MHz and 11\,MHz, respectively. The spectra are normalized to their respective maximum intensity, which is reduced by a factor of four when cooling the valve to 10\,K. The insert shows the determined Doppler-corrected transition frequencies on a magnified scale. See text for details.
\label{fig:spectra_77_10_K_Comb_Paper}}
\end{figure}

Fig.~\ref{fig:spectra_77_10_K_Comb_Paper} illustrates the $1_2\rightarrow 81$p$1_2$ transition recorded at valve temperatures of $77$\,K (red) and $10$\,K (blue). The two spectra clearly demonstrate the reduction of the Doppler width and Doppler shift associated with the reduction of the molecular-beam velocities by a factor of about two.
The geometrical centers of the two doublets are 34284.518644(23)\,cm$^{-1}$ ($77$\,K) and 34284.518657(8)\,cm$^{-1}$ ($10$\,K), which corresponds to a difference of ${\sim}400$\,kHz, in agreement with the combined experimental uncertainty of ${\sim}700$\,kHz of these values. 
The slight asymmetry in the lines that is observed in the spectrum recorded at 77\,K is a consequence of the slightly imperfect alignment of the laser and molecular beams. This asymmetry can be reduced by lowering the velocity of the molecular beam and/or improving the laser wavefront, but it does not affect the Doppler-corrected line positions because the two lines of the Doppler pair display an opposite asymmetry.

The extrapolation to the series limit relies on MQDT as developed and implemented by Ch.\,Jungen \cite{jungen1977,jungen1998,jungen2011}. The MQDT parameters of triplet $n$p and $n$f Rydberg states determined by Sprecher \textit{et al.} \cite{sprecher2014} were employed without change. In order to ensure the robust extrapolation of ionization energies, we adopt the approach of Beyer \textit{et al.} \cite{beyer2019}. The experimental data set used for the extrapolation is chosen so as to cover an energy range at least as large as the energy interval to be extrapolated.
The extrapolated ionization energy of the $N^{\prime\prime}=1$ rotational level of He$_2^*$ amounts to $\tilde{\nu}_{11}=\IPFinalValue$\,cm$^{-1}$.
The comparison between the centers of gravity of the observed transitions and calculated values resulting from the MQDT fit are also shown in Table~\ref{tab:transitions_table_Comb_Paper}.
The root mean square (RMS) of the discrepancies amounts to 740\,kHz and the residuals appear to be normally distributed.

The data also allow the determination of the first rotational interval of He${_2}^+$, provided the rotational energy corresponding to the $N^{\prime\prime}=1\rightarrow N^{\prime\prime}=3$ transition in He$_2^*$ is known with sufficient accuracy. 
We determined this value to be $\tilde{\nu}_{31}^{\prime\prime}=\LowRotIntMetastableFinalValue$\,cm$^{-1}$ by taking the difference of the $1\rightarrow 111$p$1_2$ [34292.259569(17)\,cm$^{-1}$] and $3\rightarrow 111$p$1_2$ [34216.446617(10)\,cm$^{-1}$] transition wavenumbers. We find that the interval between the first two rotational states of He${_2}^+$ is $\tilde{\nu}_{13}^+=\LowRotIntIonFinalValue$\,cm$^{-1}$.

\begin{figure}[!ht]
\centering
\includegraphics[width=.9\columnwidth]{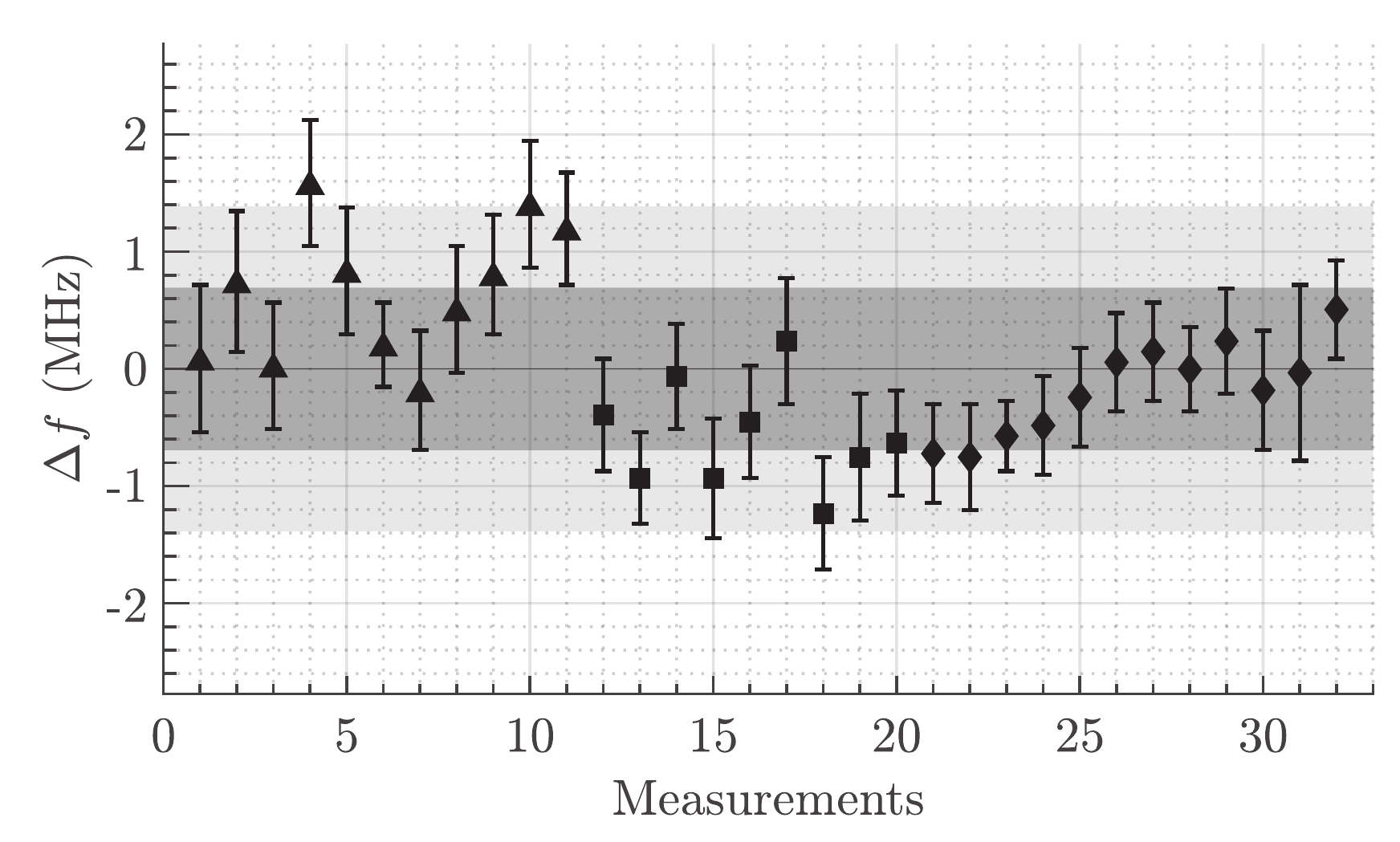}
\caption{Statistical analysis performed on the $1_1\rightarrow 97$p$1_2$ transition ($f_\mathrm{mean}/c=34289.587861(23)$\,cm$^{-1}$). The figure shows the difference $\Delta f = f-f_\mathrm{mean}$ between the transition frequency recorded on different days with the average value $f_\mathrm{mean}$ of all measurements. The different markers (triangles, squares and diamonds) denote different data sets after subsequent realignment procedures. The grey-shaded areas include the data points within one and two standard deviations, respectively ($1\sigma=700$\,kHz).
\label{fig:statistics_and_error_budget_Comb_Paper}}
\end{figure}

\begin{table}[!t]
\squeezetable
\caption{\label{tab:ErrorBudget_Comb_Paper}
Systematic and statistical ($1\sigma$, in kHz) contributions to the uncertainty in the determination of the ionization potential $\tilde{\nu}_{11}$ of He$_2^*$ (third column) and in the lowest rotational interval $\tilde{\nu}_{13}^{+}$ of He${_2}^+$ (second column). All systematic uncertainties with the exception of the MQDT-extrapolation uncertainty cancel out in the determination of the rotational interval.}
\begin{tabular*}{\columnwidth}{l@{\extracolsep{\fill}} l c c}
\toprule
\multicolumn{2}{l}{Uncertainty (kHz)}& \multicolumn{1}{l}{$\tilde{\nu}_{13}^{+}$} & \multicolumn{1}{l}{$\tilde{\nu}_{11}$}\\
\hline
\multicolumn{3}{l}{\emph{Systematic}}\\
    & ac-Stark shift 		 &	&$\ll1$ 	\\
  & photon-recoil shift &  & $\ll1$\\
  & second-order Doppler shift &		&$<1$\\ 
  & pressure shift 		&	& 5\\
  & Zeeman shift 		&	&10	\\
   & dc-Stark shift 		 &	&200 	\\
 & MQDT extrapolation &	1700	&1100\\
\hline
\multicolumn{2}{l}{\emph{Statistical\,\footnote{Including the contribution of the residual first-order Doppler shift (${\sim}\FirstOrderDopplerUncertaintyCombPaper$\,kHz) and the lineshape model (${\sim}400$\,kHz) (see text).}}} & 700 & 700 \\
\toprule
\end{tabular*}
\end{table}
The estimated statistical and systematic uncertainties affecting the recorded transition frequencies are summarized in Table~\ref{tab:ErrorBudget_Comb_Paper}.
The dominant contribution to the systematic uncertainty comes from the MQDT extrapolation. In future this value might be reduced by improving the quantum defects of Ref.~\cite{sprecher2014}, which will require accurate measurements at principal quantum numbers $n$ in the range 20--40. The largest remaining contributions to the uncertainty are the uncertainty in the lineshape model, which corresponds to $\sim$1/20 of the linewidth, the residual dc-Stark shift and the uncertainty in the first-order Doppler shift. 
The latter is estimated from the day-to-day reproducibility of the $1_1\rightarrow 97$p$1_2$ transition frequency after realignment of the laser beams through the chamber (see Fig.~\ref{fig:statistics_and_error_budget_Comb_Paper}). 
 These measurements were also used to estimate the overall statistical reproducibility to \ReproducibilityUncertaintyCombPaper\,kHz. The recoil shift and the second-order Doppler shift amount to 50\,kHz and -5\,kHz, respectively, and were compensated when deriving the final transition frequencies.
The remaining contributions to the error budget were too small to be quantified in our experiment and have been estimated. 

We improved the uncertainty of the adiabatic ionization energy of the $a\,^3\Sigma_u^+$ state of He$_2$ (\IPFinalPrecision\,MHz) and the first rotational interval of He${{_2}^+}$ (1.9\,MHz) by a factor of 30 and 10, respectively, compared to our previous studies (see Table~\ref{tab:ResultsComparison_Comb_Paper}). This improvement was achieved by replacing the pulsed UV laser used for photoexcitation by a single-mode cw UV laser and by calibrating the laser frequency with a frequency comb rather than a commercial wavemeter. Our measurements provide benchmark data that can be used to test \textit{ab initio} calculations of He${_2}^+$ and He$_2^*$ and approach the precision required to probe the polarizability of atomic helium via the two-body interaction potential in He${_2}^+$. Our results are relevant in the context of the establishment of primary pressure standards and complement the current most accurate results that rely on dielectric-constant gas thermometry \cite{gaiser2018}.

\begin{acknowledgments}
We are grateful to Ch. Jungen for letting us use his MQDT program and to Josef Agner for excellent technical support. We thank B. Jeziorski (Warsaw), E. M\'{a}tyus (Budapest), N. H\"{o}lsch (Zurich) and M. Beyer (Yale) for useful discussions. This work was supported by the Swiss National Science Foundation (Grant No. 200020-172620) and by the European Research Council (Horizon 2020, Advanced Grant 743121).
\end{acknowledgments}
\begin{table}[!ht]
\squeezetable
\caption{\label{tab:ResultsComparison_Comb_Paper}
Comparison of the adiabatic ionization energy of He$_2^*$ ($\tilde{\nu}_{11}$) and of the lowest rotational interval of He$_2^*$ ($\tilde{\nu}_{31}^{\prime\prime}$) and of He${_2}^+$ ($\tilde{\nu}_{13}^+$) obtained in this work and in previous studies. All values are given in cm$^{-1}$.
}
\begin{tabular*}{\columnwidth}{
l@{\extracolsep{\fill}} 
S[table-format=4.5]
S[table-format=3.7]
S[table-format=3.7]
}
\hline\hline
\multicolumn{1}{c}{Reference}&  \multicolumn{1}{c}{$\tilde{\nu}_{11}$} & \multicolumn{1}{c}{$\tilde{\nu}_{31}^{\prime\prime}$} & \multicolumn{1}{c}{$\tilde{\nu}_{13}^+$}\\
\hline
Ref. \cite{ginter1980} &\IPGinterJMS	&	& \LowRotIntIonGinterJMS	\\
Ref. \cite{focsa1998} &	&	\LowRotIntMetastableFocsaJMS	&	\\
Ref. \cite{raunhardt2008}	& \IPRaunhardtJCP & \LowRotIntMetastableRaunhardtJCP	&\LowRotIntIonRaunhardtJCP \\
Ref. \cite{tung2012} 	&	&	& \LowRotIntIonTungJCP	\\
Ref. \cite{sprecher2014}	&\IPSprecherJCP &	&\LowRotIntIonSprecherJCP	\\
Ref. \cite{jansen2015}	& \IPJansenPRL	&	\LowRotIntMetastableJansenPRL&\LowRotIntIonJansenPRL	\\
Ref. 	\cite{matyus2018}	&	&	&\LowRotIntIonMatyusJCP \hspace{-0.75cm} \footnote{Containing an estimated value of the relativistic and QED corrections. The nonadiabatic term value is 70.936\,cm$^{-1}$ \cite{matyus2018}.}	\\
This work & \IPFinalValueAllUncertainties	& 75.812953(20) &\LowRotIntIonFinalValueAllUncertainties	\\
\hline\hline
\end{tabular*}
\end{table}


%

\end{document}